\documentclass[prl,reprint,superscriptaddress,showpacs]{revtex4-1}
\usepackage{graphicx}
\usepackage[utf8]{inputenc}
\usepackage{amsmath}
\usepackage{dcolumn}

\begin{document}

\title{Non-Gaussian Current Fluctuations in a Short Diffusive Conductor}

\author{Edouard Pinsolle}
\author{Samuel Houle}
\author{Christian Lupien}
\author{Bertrand Reulet}
\affiliation{Institut Quantique, D\'{e}partement de Physique, Universit\'{e} de Sherbrooke, Sherbrooke, Qu\'{e}bec J1K 2R1, Canada.}
\date{\today}

\begin{abstract}
	We report the measurement of the third moment of current fluctuations in a short metallic wire at low temperature. The data are deduced from the statistics of voltage fluctuations across the conductor using a careful determination of environmental contributions. Our results at low bias agree very well with theoretical predictions for coherent transport with no fitting parameter. By increasing the bias voltage we explore the cross-over from elastic to inelastic transport.  
\end{abstract}

\maketitle
Over the years the study of current fluctuations, and in particular their variance (or second moment) $\langle\delta I^2\rangle$, has given new insights in the properties of quasi-particles in conductors. The study of the symmetry of the Kondo state in carbon nanotubes \cite{ferrier_universality_2015} or the observation of heat quantization \cite{jezouin_quantum_2013} are recent examples of such measurements.
Despite this success there has been only a few attempts to push deeper the study of current fluctuations in mesoscopic conductors by tackling the measurement of higher order moments such as the third one $\langle\delta I^3\rangle$. As far as we know, such measurements have been performed only in tunnel junctions \cite{reulet_environmental_2003,bomze_measurement_2005,gabelli_dynamics_2008,gabelli_high_2009}, quantum dots \cite{gustavsson_counting_2006, gustavsson_electron_2009, sukhorukov_conditional_2007, gustavsson_measurements_2007} and avalanche diodes \cite{Gabelli_full_2009}, revealing physics hidden in the study of the second moment such as the coupling with the environment \cite{reulet_environmental_2003}, the dynamics of quantum noise \cite{gabelli_dynamics_2008} or the ordering of operators in a quantum measurement \cite{gabelli_high_2009}. 
Higher order moments of current fluctuations in even the simplest system, an electrical wire, has never been probed experimentally. This letter reports such a measurement.

The variance of current fluctuations in a diffusive wire has been calculated using many techniques \cite{beenakker_suppression_1992,nagaev_shot_1992,roche_mesoscopic_nodate}, all providing the same answer for the spectral density of current fluctuations $S_{I^2}$ measured at temperature $T$ with a voltage bias $V$:

\begin{equation}
	S_{I^2}=\frac{2}{3}\frac{2k_BT}{R}+\frac{1}{3}\frac{eV}{R}\coth{\frac{eV}{2k_BT}},
  \label{eq_SI2}
\end{equation}
\noindent where $R$ is the sample resistance, $e$ the electron charge and $k_B$ the Boltzmann's constant. This result indicates the existence of shot noise with a Fano factor $F_2=e^{-1}dS_{I^2}/dI=1/3$ at large bias $V\gg k_BT/e$, which has been confirmed experimentally \cite{henny_1/3-shot-noise_1999}. The reduction of the Fano factor as compared to that of a tunnel junction, $F_2=1$, is interpreted in the quantum theory as stemming from the existence of well transmitting channels and in the semi-classical theory from the existence of a position-dependent distribution function. The third moment of current fluctuations has also been calculated by several theories \cite{nagaev_cascade_2002,salo_frequency-dependent_2006,roche_mesoscopic_nodate,gutman_shot_2003,roche_shot-noise_nodate,lee_universal_1995} which at low frequency all yield to the same spectral density $S_{I^3}$ given by:

\begin{equation}
	S_{I^3}=\frac{1}{15}e^2I+\frac{12}{5}k_BT \frac{dS_{I^2}}{dV}.
	\label{eq_SI3}
\end{equation}
\noindent This result differs from that of a tunnel junction $S_{I^3}=e^2 I$ on two main factors: first it depends on temperature; second it has a much lower Fano factor at high voltage, $F_3=e^{-2}dS_{I^3}/dI=1/15$ instead of $F_3=1$ for the tunnel junction. While this prediction comes in the quantum regime again from the statistical distribution of transmissions, the semi-classical prediction involves a ``cascade" or feedback mechanism similar to that explaining environmental contributions.

Eq. (\ref{eq_SI3}) corresponds to a measurement performed with a noiseless voltage bias and an ammeter, i.e. an apparatus with an input impedance much lower than that of the sample. This situation can be achieved with a high impedance sample, but a typical metallic wire has a low impedance and one has to consider the effects of both the finite impedance of the environment, here a resistance $R_A$ (the input impedance of the amplifier), and the current noise experienced by the sample, here generated by the amplifier used to detect current fluctuations and described by a noise spectral density $S_A$. These contributions to the spectral density of the variance of \emph{voltage} fluctuations across the amplifier's input resistance are simply given by:

\begin{equation}
  S_{V^2}=R_{D}^2 \left(S_{I^2}+S_A \right),
	\label{eq_SV2}
\end{equation}
\noindent with $R_D=RR_A/(R+R_A)$. Here the environment only renormalizes the noise generated by the sample and adds a contribution which does not depend on the temperature or bias voltage. In contrast, environmental effects are much more subtle on the third moment of voltage fluctuations. They have been thoroughly studied both theoretically \cite{beenakker_temperature-dependent_2003,kindermann_feedback_2004} and experimentally \cite{reulet_environmental_2003,bomze_measurement_2005} and obey: 

\begin{equation}
	S_{V^3}=-R_{D}^3S_{I^3}+3R_{D}^4(S_A+S_{I^2})\frac{dS_{I^2}}{dV}.
  \label{eq_env}
\end{equation}
\noindent As a consequence, a reliable way to characterize all the environmental terms is required to extract the intrinsic third moment of current fluctuations $S_{I^3}$. 

In the following we show the measurement of $S_{I^3}$ for a short metallic wire placed at very low temperature. We describe the experimental setup and the results for $S_{V^3}$, the calibration of the environmental contributions using a tunnel junction and the results for $S_{I^3}$. We compare these results with the theoretical prediction of Eq. (\ref{eq_SI3}) in the elastic transport regime. We also report measurements performed at high bias that explore the crossover to the inelastic regime.

\begin{figure}
\center
	\includegraphics[width=1\columnwidth]{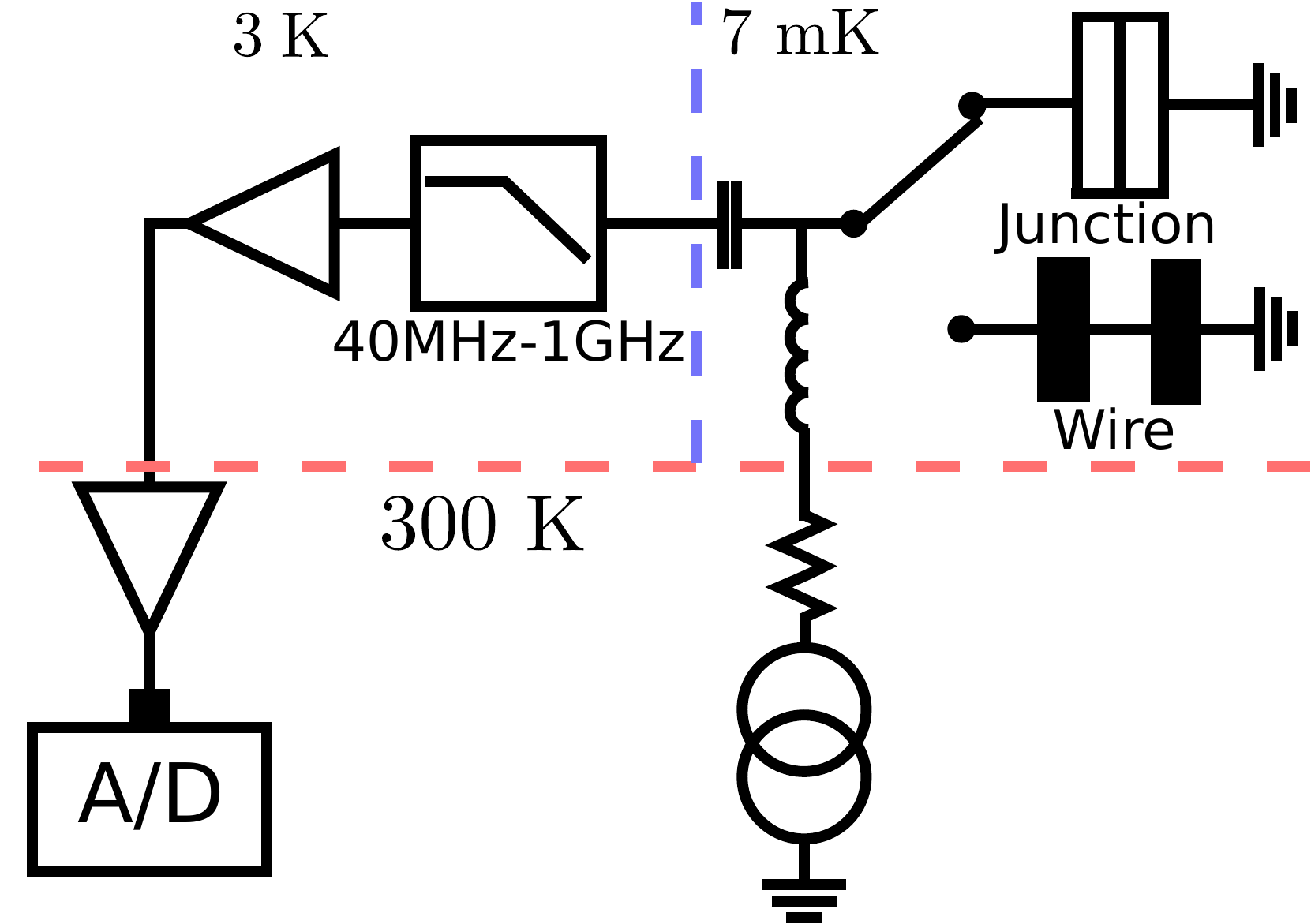}
	\caption{Schematics of the experimental setup. A/D represents a 14 bits, 400 MSample/s digitizer.}
\label{schema}
\end{figure}

\emph{Experimental setup.} The sample is a $1$ $\mu$m long, 10 nm wide, 165 nm thick Aluminum (Al) wire of resistance $R_{w}=30.5$ $\Omega$. Its contacts, also made of Al, are much larger ($400$ $\mu$m$\times400$ $\mu$m) and thicker ($200$ nm) to make sure they behave as good electron reservoirs \cite{giazotto_opportunities_2006}.
An Al tunnel junction of resistance $R_j=34$ $\Omega$ is used as a reference to calibrate the setup. 
Both samples have been made by e-beam lithography and the metal has been deposited by double angle evaporation \cite{reese_niobium_2007}.
The experimental setup is presented in Fig. \ref{schema}. 
The samples are placed on the 7 mK stage of a dilution refrigerator. They are kept in their normal, non superconducting state with the help of a strong Neodymium permanent magnet. The two samples are connected to a cryogenic microwave switch which allows us to measure either of them without changing anything in the detection circuit.  
They are dc current biased through the dc port of a bias-tee and ac coupled to a cryogenic microwave amplifier in the range $40$ MHz-$1$ GHz. The use of a cryogenic amplifier both optimizes the signal to noise ratio and minimizes the noise experienced by the sample which leads to environmental contributions. The signal is further amplified at room temperature in order to achieve a level high enough for digitization. Non-linearities in the detection are very detrimental since they lead to strong artifacts. Despite the use of ultra-linear amplifiers, non-linearities still give rise to a contribution which is an even function of $I$ in the sample. We simply remove this by considering $[S_{V^3}(I)-S_{V^3}(-I)]/2$. After amplification the signal is digitized by a 14 bit, 400 MS/s digitizer with a 1 GHz analog bandwidth. We measure real-time histograms of the signal from which moments are computed.

\begin{figure}
\center
	\includegraphics[width=1\columnwidth]{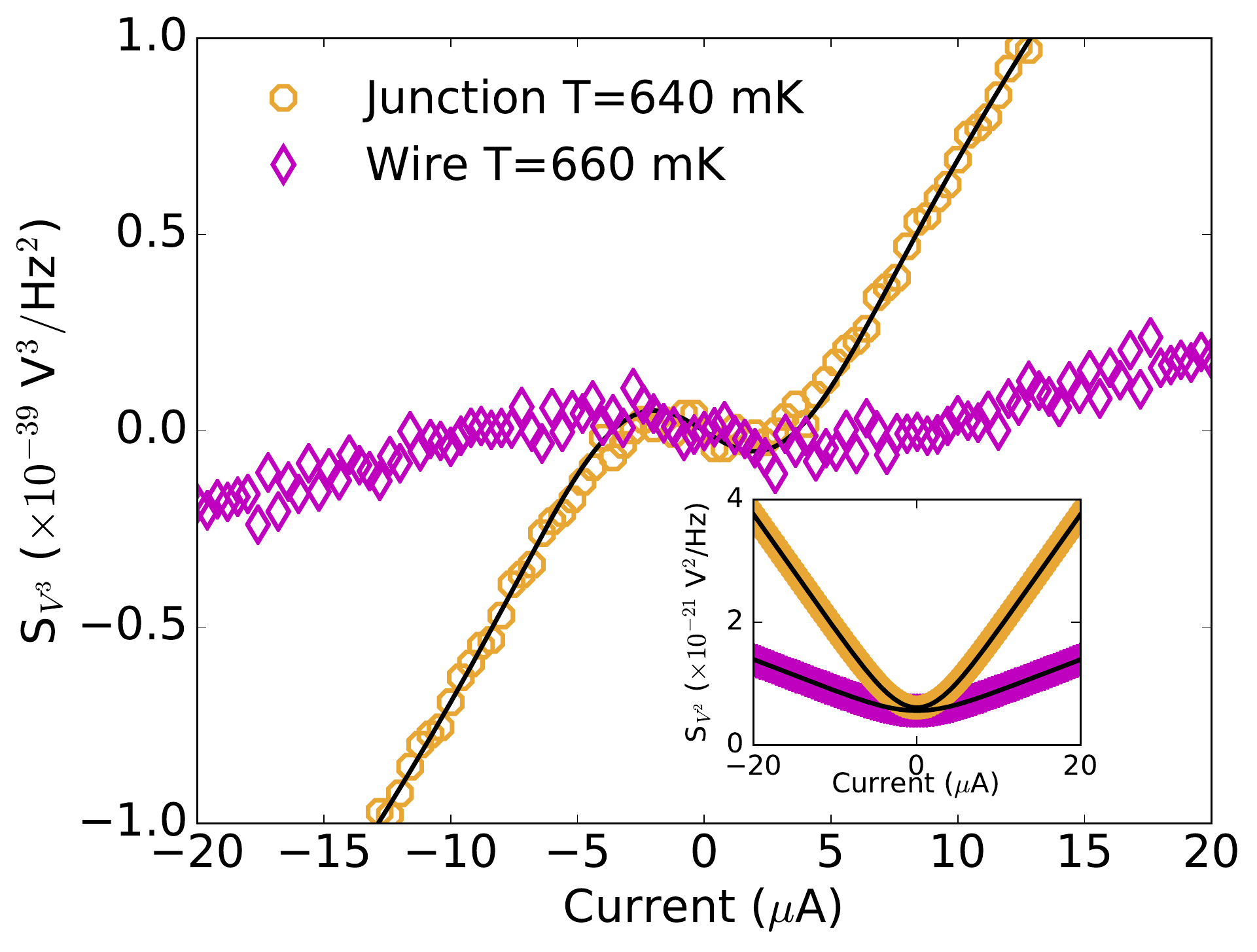}
	\caption{Third moment of voltage fluctuations $S_{V^3}$ for the wire (purple) and the tunnel junction (orange) vs. $I$. Symbols are experimental data; the solid line is a fit for the tunnel junction using Eq. (\ref{eq_env}) with $S_{I^3}=e^2I$, from which the environmental parameters are deduced. Inset: the second moment of current fluctuations $S_{I^2}$ vs. $I$ (the noise of the amplifier has been subtracted for the sake of clarity).}
\label{S3_comp}
\end{figure}

\begin{figure}
\center
	\includegraphics[width=1\columnwidth]{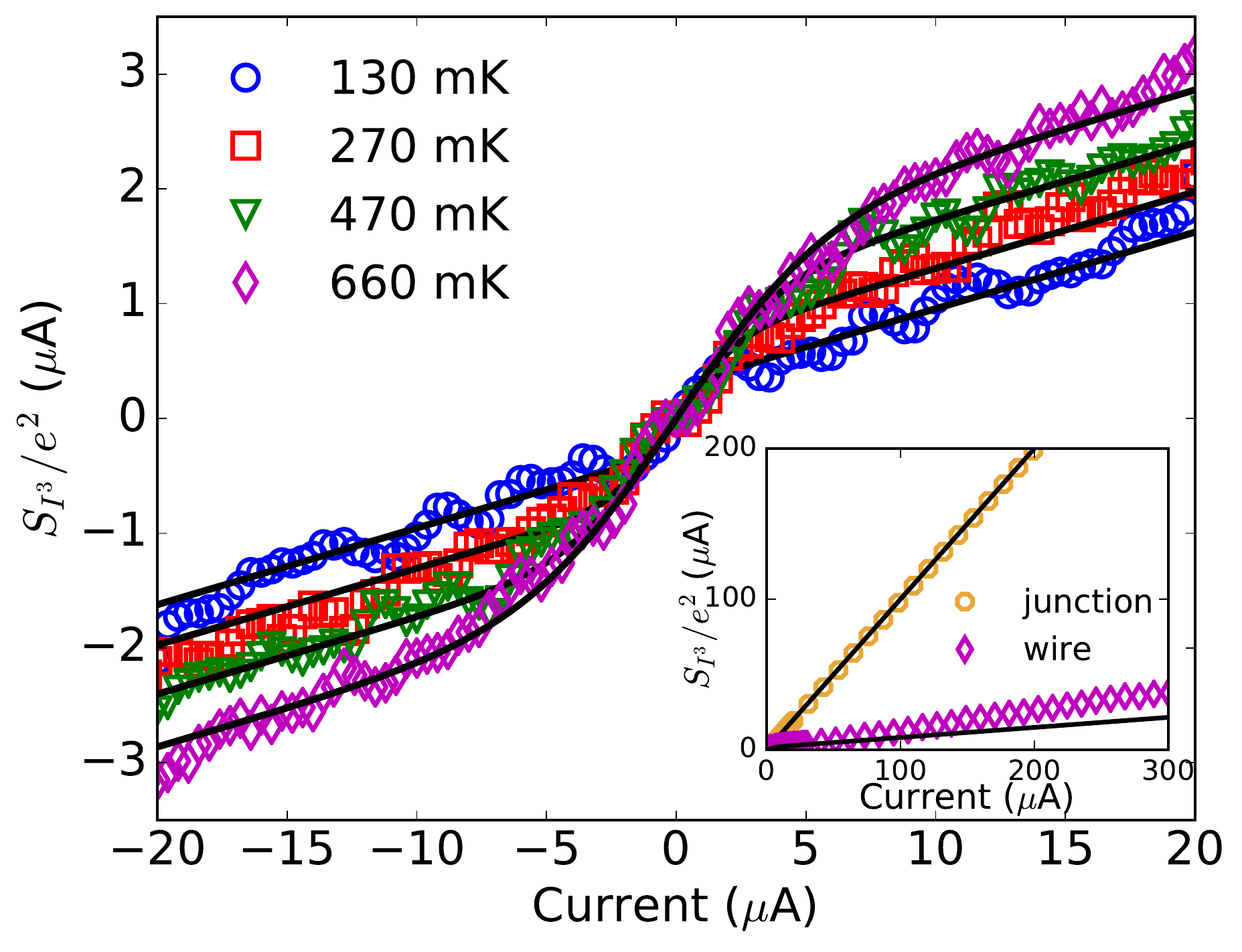}
	\caption{Intrinsic third moment of current fluctuations $S_{I^3}$ vs. $I$ for the wire at different temperatures. Symbols are  experimental data, lines are the theoretical expectations of Eq. (\ref{eq_SI3}). Inset: $S_{I^3}$ for the tunnel junction (orange) and the wire (purple) for higher currents up to 0.3 mA at $T\sim640$~mK.}
\label{SI3}
\end{figure}

\emph{Results: elastic transport.} In the inset of Fig. \ref{S3_comp} we present the measurement of $S_{V^2}$ for the tunnel junction (orange symbols) and the wire (purple). From the high current slope of $S_{V^2}$ vs. $I$ for the tunnel junction we find the gain of the setup. Then, we deduce the Fano factor of the wire $F_2=0.35\pm0.02$, in good agreement with the theoretical value of $\frac{1}{3}$ in Eq. (\ref{eq_SI2}). This ensures that electron transport is elastic in the sample, in agreement with other measurements of similar wires \cite{pothier_energy_1997,henny_1/3-shot-noise_1999}. From $S_{V^2}$ we also deduce the electron temperature for the wire and for the tunnel junction, as well as the noise temperature of the amplifier $T_a\simeq7.5$ K. Values of the temperature indicated in the various figures correspond to electronic temperatures deduced from the measurements of $S_{V^2}$.\
Fig. \ref{S3_comp} presents the measurement of $S_{V^3}$ vs. $I$ for the tunnel junction and the wire at a temperature around $650$ mK (we have performed similar measurements down to $120$ mK). As a consequence of small Fano factors $F_2$ and $F_3$ the signal is much weaker for the wire, but the signal-to-noise ratio clearly allows for a reliable comparison with theory (each point is averaged for 20 min). Following the procedure of \cite{gabelli_electronphoton_2013}, we use the measurements performed at all temperatures on the tunnel junction to extract the parameters that characterize the environment, i.e. the amplifier impedance $R_A=44.8$ $\Omega$ and the effective environmental noise temperature $T_0^*=0.54$ K. An example of fit on the tunnel junction $S_{V^3}$ is provided as a solid line on Fig. \ref{S3_comp}. From the knowledge of the environmental parameters we can extract the intrinsic third moment of current fluctuations in the wire using Eq. (\ref{eq_env}). The corresponding results are plotted in Fig. \ref{SI3}. The theoretical predictions of Eq. (\ref{eq_SI3}) are plotted as solid lines with no fitting parameters. A clear agreement between experiment and theory is achieved at all temperatures for the current range explored. At low current ($eV<k_BT$) we observe that all curves merge, which corresponds to a Fano factor $F_3=1/3$ independent of temperature (the fact that the Fano factor of $S_{I^3}$ at low current is the same as that of $S_{I^2}$ at high current has been predicted to come from the Pauli principle \cite{roche_shot-noise_nodate,roche_mesoscopic_nodate}). At high current, we observe that $S_{I^3}$ grows linearly with current with a slope $F_3=1/15$ that is temperature independent. However there is a constant shift which increases with temperature. In this limit, both $S_{I^2}$ and $S_{I^3}$ increase linearly with temperature, while for a tunnel junction none of them do.

\begin{figure}
\center
	\includegraphics[width=1\columnwidth]{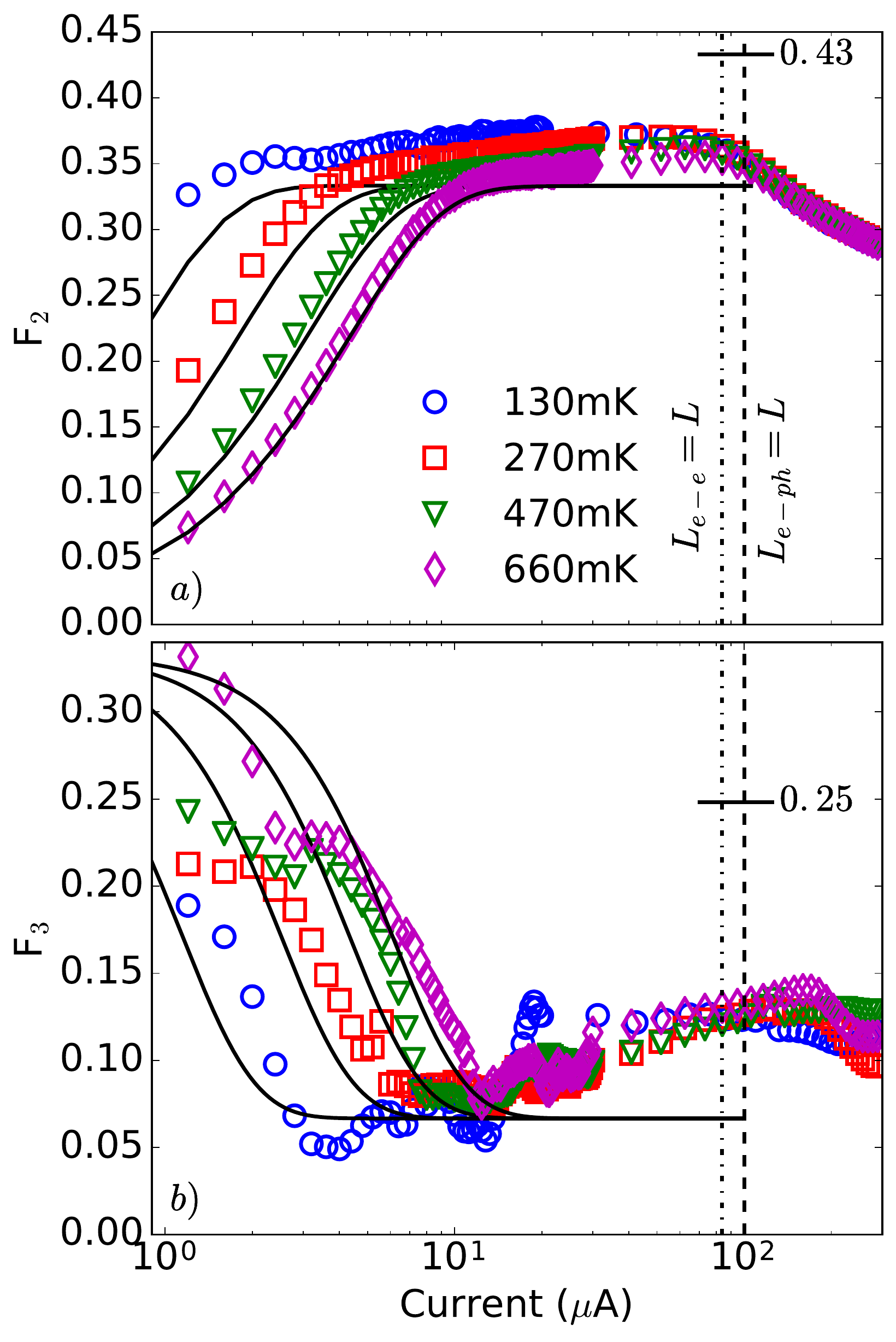}
	\caption{a) Fano factor of the second moment of current fluctuation $F_2=e^{-1} dS_{I^2}/dI$ for the wire vs. $I$. b) Fano factor of the third moment of current fluctuations $F_3=e^{-2}dS_{I^3}/dI$ for the wire vs. $I$. Different symbols correspond to different temperatures ranging from 130 mK to 660 mK. The black lines correspond to the theoretical prediction in the elastic regime. Dashed (dotted dashed) line corresponds to the current for which $L_{e-e}=L$ ($L_{e-ph}=L$). }
\label{F3}
\end{figure}

\emph{From elastic to inelastic transport.} Electronic transport in short samples at very low energy (voltage and temperature) is elastic. When energy is increased, inelastic processes are more and more effective, starting with electron-electron interactions which tend to thermalize the electrons among themselves, followed by electron-phonon interactions which tend to thermalize the electrons to the phonon bath of the substrate.
This manifests itself in the variance of current fluctuations in a wire by the Fano factor $F_2$ going from $1/3\simeq0.33$ in the elastic regime to $\sqrt{3}/4\simeq0.43$ in the hot electron regime (strong electron-electron, no electron-phonon interaction) and decaying to zero as the electron-phonon interaction becomes effective \cite{steinbach_observation_1996}. We show in Fig. \ref{F3}(a) a lin-log plot of $F_2$ for our wire vs $I$. $F_2$ is obtained by taking the numerical derivative of $S_{I^2}$ vs. $I$ after smoothing of the experimental data. After a sharp rise when $eV<k_BT$, $F_2$ has a short plateau at $\sim0.35$. This corresponds to the elastic regime described by Eq. (\ref{eq_SI2}), shown as solid lines in the figure. At higher bias $F_2$ slowly increases up to $\sim0.37$ at $I\sim80$~$\mu$A, followed by a decay down to $0.28$ at $I=0.3$ mA. We could not apply a stronger current in the sample without taking the risk to damage it. We do not observe a plateau at $F_2\simeq0.43$ that would correspond to inelastic scattering being dominated by electron-electron interactions.

Inelastic processes should also affect the third moment of current fluctuations. We show in Fig. \ref{F3}(b) the Fano factor $F_3$ vs. $I$, deduced from $S_{I^3}$. The theoretical prediction of Eq. (\ref{eq_SI3}) (elastic transport) corresponds to $F_3$ going from $1/3$ at $eV<k_BT$ to $1/15$, as shown as a solid line in Fig. \ref{F3}(b). We clearly observe this transition in the experimental data, even though we explore the regime $eV<k_BT$ deep enough only at the highest temperature, where the plateau $F_3=1/3$ is visible. $F_3$ in the hot electron regime has been predicted to be given by $8/\pi^2 -9/16\simeq0.25$ while electron-phonon interaction is expected to lead to a vanishing $F_3$ at high enough energy. We indeed observe an increase of $F_3$ up to $\sim0.14$ at $I\sim150$ $\mu$A followed by a slow decrease down to $F_3\sim 0.10$ at $I=0.3$ mA.

The crossover between the different regimes should occur when the length of the sample is of the order of the corresponding inelastic length, $L_{e-e}$ for electron-electron interaction or $L_{e-ph}$ for electron-phonon interaction. Both decrease when the bias voltage or the electron temperature is increased. Thus an increase of the current flowing through the sample tends to decrease the inelastic lengths. Two vertical lines in Fig. \ref{F3} represent the calculated current for which $L=L_{e-e}$ (left) and $L=L_{e-ph}$ (right). Hence the left part of the plots corresponds to the elastic regime ($L>L_{e-e},L_{e-ph}$), the region between the vertical lines correspond to the hot electron regime ($L_{e-e}<L<L_{e-ph}$) and the right region to the phonon cooled regime ($L>L_{e-e},L_{e-ph}$). The intermediate region where electron-electron interaction dominates is clearly very narrow, which explains why we never observe the expected values of the Fano factors calculated in the hot electron regime. To our knowledge the crossover between the elastic and electron-electron regimes has never been theoretically studied. 
Deep in the electron-phonon regime, the noise in a wire is understood as Johnson-Nyquist noise of a sample at a uniform electronic temperature $T$ determined by the balance between Joule heating and phonon cooling, i.e.:
\begin{equation}
RI^2=\Sigma \Omega \left(T^n - T_{ph}^n\right),
\label{eq_eph}
\end{equation}
\noindent where $T_{ph}$ is the phonon temperature, $\Sigma$ the electron phonon coupling constant, $\Omega$ the sample volume and n a number ranging from 4 to 6 depending on the sample purity. Thus the noise spectral density $S_{I^2}=2k_BT/R$ increases with current with a decaying Fano factor $F_2\propto I^{2/n-1}$ at large current. However $S_{I^3}$ in the electron-phonon regime has been calculated to decrease to zero at high current (or zero phonon temperature $T_{ph}=0$) as \cite{gutman_shot_2003,huard_electron_2007}:
\begin{equation}
	S_{I^3}\propto\frac{k_B^2R^{2/n-2}}{(\Sigma \Omega) ^{2/n}}I^{4/n-1}.
\label{eq_S3eph}
\end{equation}
This result can be understood using the following simple model, which gives the same result as a full calculation  using cascaded Boltzmann-Langevin equations with a position-dependent distribution function for the electrons. Let us consider the wire of length $L\gg L_{e-ph}$ as many wires connected in series. Each wire exhibits thermal noise with zero third moment. However for each wire, all the others play the role of an electromagnetic environment that has voltage-dependent noise, i.e. leads to environmental contributions. This immediately leads to:
\begin{equation}
S_{I^3}=3S_{I^2}\frac{d S_{I^2}}{d I},
\end{equation}
which gives Eq. (\ref{eq_S3eph}) for Johnson-Nyquist noise at a temperature T given by Eq. (\ref{eq_eph}). In contrast with $S_{I^2}$, $S_{I^3}$ is expected to exhibit a maximum in current, then a decay at high current. 
We show in the inset of Fig. \ref{SI3} our measurements of $S_{I^3}$ vs. $I$ up to $0.3$ mA. We observe that $S_{I^3}$ for the tunnel junction is strictly linear up to the highest current, as expected. The wire deviates from the linear behavior of Eq. (\ref{eq_SI3}) and barely shows any sign of saturation (but $F_3$ decays slightly for the highest current, see Fig. \ref{F3}(b)). For this sample $\Sigma$ and $n$ have been measured \cite{pinsolle_direct_2016} and we expect to observe $F_3\propto I^{-6/5}$ at a current of $I\gtrsim 0.5$ mA, which we did not reach.

It is noteworthy that taking $n=4$ in Eq. (\ref{eq_eph}) leads to a third moment of current fluctuations saturating at high current, and not decaying. Such a value of $n$ has been observed in thin Au wires \cite{dorozhkin_energy_1986}. A saturation of $S_{I^3}$ would however not violate the central limit theorem, neither when considering a sample of arbitrary length or an arbitrary large current since $S_{I^3}/S_{I^2}^{3/2}\propto I^{-3/4}L^{-1/2}$.

\emph{Conclusion.}
We have measured the third moment of current fluctuations in a wire, thus demonstrating that even the simplest conductor exhibits non-Gaussian noise. Our data at low voltage are in very good quantitative agreement with the theory. In particular we have found a Fano factor $F_3=1/15$ characteristic of elastic transport in diffusive conductors. At higher current we recorded a clear deviation from the elastic regime, when inelastic scattering lengths become of the order of the sample length.\\
Third moment of current fluctuations, whether intrinsic or environmentally induced, have been shown to affect decoherence in quantum dots \cite{galperin_non-gaussian_2006} and should be taken into account to explain observed coherence times.\\  
Our experiment also demonstrates the possibility to measure the third moment of current fluctuations with a great bandwidth $\sim 1$GHz (and thus a great sensitivity) in conductors of moderate resistance, while also getting rid of environmental contributions thanks to a cryogenic calibration procedure. This opens the possibility to study many more systems where the statistics of electron transport is more complex. For example in the presence of proximity effect where multiple Andreev reflections have been predicted to lead to a diverging third moment \cite{Pilgram_Noise_2005,cuevas_full_2003}, in the vicinity of a phase transition \cite{bagrets_nonequilibrium_2014} or in conductors where electron-electron interactions are more prominent. 

We acknowledge technical help of G. Lalibert\'{e}.
This work was supported by the Canada Excellence Research Chair program, the NSERC the MDEIE, the FRQMT via the INTRIQ, the Universit\'{e} de Sherbrooke via the EPIQ and the Canada Foundation for Innovation.

\bibliographystyle{apsrev4-1}
\bibliography{S3_library}

\end{document}